\documentclass[preprint,12pt]{elsarticle}

\usepackage[T1]{fontenc}
\usepackage[utf8]{inputenc}
\usepackage{graphicx}
\usepackage{booktabs}
\usepackage{amsmath}
\usepackage{listings}
\usepackage{xcolor}
\usepackage{float}
\usepackage{url}
\usepackage{placeins}

\newfloat{listing}{htbp}{lol}
\floatname{listing}{Listing}

\journal{Computers \& Security}

\begin{document}

\begin{frontmatter}

\title{The Exclusion Ratchet: False-Positive Suppression Accumulates and
       Persists in Detection Rule Repositories}

\author[ind]{Sudaroli Dhananjeyan\corref{cor}}
\ead{oli.sudar@gmail.com}
\cortext[cor]{Corresponding author}

\author[amrita]{Kumaran U}
\ead{u_kumaran@blr.amrita.edu}

\affiliation[ind]{organization={Independent Researcher},
                  city={Bengaluru},
                  country={India}}

\affiliation[amrita]{organization={Department of Computer Science and
                     Engineering, Amrita School of Computing, Amrita Vishwa
                     Vidyapeetham},
                     city={Bengaluru},
                     country={India}}

\begin{abstract}
When a rule produces too many false alarms an analyst adds an exclusion, and
the rule thereafter declines to alert in that circumstance. Each such
decision is locally reasonable; what becomes of them collectively is not
known. Recent longitudinal work established that curation does not converge,
but measured restoration time only for revisions that were later reverted
--- a measure silent about narrowing that is never undone. We measure that.
Across nine years and 8,234 revisions of the SigmaHQ corpus we detect
suppression semantically --- growth in the set of predicates held under
negation without compensating growth in coverage --- and validate it against
blinded hand labelling (precision 0.828, recall 0.911). The test is
deterministic: nothing is learned from the data, and the definitions are
released as code. Exclusions were added 1,642 times and withdrawn 304, a
ratio of 5.4 to 1 that rises to 13 to 1 at the level of the individual rule.
Thirty-one per cent of the narrowing is invisible to structural comparison,
which existing structural accounts therefore undercount. Estimated by
Kaplan--Meier, 86.7 per cent of exclusions remain in force three years on,
and persistence is independent of whether the rule is the only coverage for
its ATT\&CK technique ($p = 0.49$). Of path-valued exclusions, 64.1 per cent
can be satisfied by an unprivileged process that chooses a filename.
Narrowing accumulates, is rarely revisited, and is not triaged by
consequence. We give a criterion for deciding which exclusions to examine
first.
\end{abstract}


\end{frontmatter}

\section{Introduction}
\label{sec:intro}

\subsection{The tuning problem}

A detection rule that fires too often is worse than one that fires rarely. Analysts cannot sustain attention on a rule whose output is mostly noise. In one measured security operations centre, between 24,000 and 134,000 alerts arrived per day, of which 0.01 per cent corresponded to true attacks and 49 per cent were correct matches on activity that turned out to have a business-justified explanation \citep{yang2024}; analysts themselves describe false-positive rates approaching 99 per cent \citep{alahmadi2022}. The operational response is well understood, and practitioners describe it consistently: narrow the rule \citep{vermeer2023}. Add an exclusion for the deployment tool that legitimately runs encoded PowerShell. Filter the backup process that legitimately accesses the volume directly. Each of these decisions is locally correct, made under real pressure, by someone who knows the environment.

Each also narrows what the rule can see, and the cost of that narrowing depends on something the decision itself does not record: whether an adversary could arrange to satisfy the exclusion. An exclusion written on a protected system path costs an attacker the privilege required to write there. One written on a bare filename cost nothing at all --- the attacker names a file. In the corpus examined here, 64.1 per cent of path-valued exclusions fall into the second class and 33.0 per cent into the first. The trade is real, it is usually worth making at the moment it is made, and its price varies by a factor that nobody records. That the resulting gaps are exploitable is not hypothetical: \citet{uetz2024} constructed evasions for widely deployed SIEM rules and defeated nearly half of them in a live enterprise network.

What is not known is what happens afterwards.

\subsection{What prior work established, and what it could not see}

\citet{long2026} provided the first longitudinal analysis of log-based detection-rule evolution, examining nine years of history across the Sigma and Splunk Security Content repositories. They showed that rule curation does not converge: 56 per cent of rules undergo revision to their detection logic, a majority both gain and lose conditions over their lifetime, and roughly a third alternate between expanding coverage and suppressing false positives without settling. For revisions that were reverted, they report a median restoration time of 114.5 hours.

That figure is conditioned on reversion. A rule narrowed in 2021 and still narrowed at the observation date contributes nothing to it; the population is defined by having returned. The authors are explicit that their analysis is structural rather than semantic, defining their patterns over syntactic transformations of predicate logic rather than their semantic effect on matched events. That restriction is principled, and this paper begins by showing that it is also load-bearing.

\subsection{Why structural analysis cannot locate suppression}

We first attempted to identify narrowing structurally, resolving the root operator per rule and classifying predicate additions and removals by the scope in which they occurred. The classifier was careful and it was wrong. Of the events it identified on singly-covered techniques, the majority proved on inspection to be maintenance: a single pull request removing deprecated hash-field syntax across the ruleset, sub-technique retagging, dead-code removal, a file rename.

The reason is structural. In Sigma, suppression is expressed as a negation --- the rule fires on a selection and not a filter --- or as an additional value appended to an existing filter list. Both defeat a classifier operating on the shape of the predicate graph: the first inverts the polarity of every operation inside it, and the second changes no structure at all. Of the 8,234 predicate-changing revisions in this corpus, 3,231 carry a predicate under negation and 4,174 involve a predicate-value update; 5,927, or 72 per cent, involve at least one of the two. Structural methods handle both poorly, and what remains after they are set aside is largely housekeeping.

The contribution that follows from this is not that structural analysis fails. It is that the narrowing it cannot see can be measured by other means, and that once measured it turns out to be 31 per cent of the total.

\subsection{Contributions}

This paper makes six contributions.

\begin{itemize}
  \item A deterministic semantic test for suppression in detection rules, defined over the set of predicates held under negation rather than over rule structure, with the definitions released as code and validated against blinded hand labelling at precision 0.828 (95\% CI [0.711, 0.904]) and recall 0.911.
  \item A demonstration that structural comparison undercounts narrowing by 31 per cent, because a third of it is expressed by appending a value to a list that already exists. This bears on any measurement of rule evolution, in any format with list-valued fields, and not only on Sigma.
  \item The exclusion ratchet: across nine years, exclusions are added 5.4 times for every occasion on which one is withdrawn, and 13 times for every occasion at the level of the individual rule, with 86 per cent of rules that ever modified their exclusion set ending net-narrower than they began.
  \item A survival analysis of exclusion persistence showing that 86.7 per cent remain in force three years after they are added, together with a null result: persistence is independent of whether the rule is the only coverage in the corpus for its ATT\&CK technique.
  \item A characterisation of what exclusions are written on, distinguishing those an unprivileged adversary can satisfy from those that cost privilege, and crossing that with breadth to give a criterion for which exclusions merit review first.
  \item A reproduction of the prior work performed before any extension, with four discrepancies documented and reported to its authors, and public release of the analysis pipeline, the derived data and the complete validation record.
\end{itemize}

\section{Background}
\label{sec:background}

This section gives the reader an understanding of the existence and usage of detection rules. Since this work analyses detection rules written in the Sigma format, it sets out how a Sigma rule is composed, how an exclusion is expressed within one, and how rules carry ATT\&CK tags.

\subsection{Detection rules and the Sigma format}

Security telemetry arrives at a volume no analyst can review directly, so review is mediated by detection rules. A detection rule is a standing query: a pattern over the fields of a log record, evaluated against every record in a stream of endpoint and server telemetry --- process creation, logon activity, file and registry writes. A record matching the pattern raises an alert in an analyst's queue for triage, a queue that in a measured security operations centre runs to tens of thousands of alerts a day \citep{yang2024}. Sigma is a YAML format for writing such patterns independently of any one SIEM's query language; a rule is not itself executable, but is compiled to SPL, KQL, EQL or another dialect before it runs \citep{sigmahq2024}.

Listing 1 gives one rule from the corpus, with its parts keyed to the notes beneath it. A predicate is a single field--value test --- Image|endswith: '\textbackslash{}certutil.exe' is one --- and is the unit in which this paper counts.

\begin{listing}[htbp]
\caption{Rule proc\_creation\_win\_certutil\_encode.yml (UUID e62a9f0c-ca1e-46b2-85d5-a6da77f86d1a) at the snapshot. The tags, logsource and detection blocks are verbatim; metadata fields (status, references, author, date, modified) are omitted and two descriptive strings truncated. Callouts are keyed below.}
\label{listing:1}
\begin{lstlisting}
title: File Encoded To Base64 Via Certutil.EXE
id: e62a9f0c-ca1e-46b2-85d5-a6da77f86d1a	(1)
description: Detects the execution of certutil with the
  "encode" flag to encode a file to base64. This can be
  abused for data exfiltration
tags:	(2)
    - attack.defense-evasion
    - attack.t1027
logsource:	(3)
    category: process_creation
    product: windows
detection:	(4)
    selection_img:
        - Image|endswith: '\certutil.exe'
        - OriginalFileName: 'CertUtil.exe'
    selection_cli:
        CommandLine|contains|windash: '-encode'
    condition: all of selection_*
falsepositives:
    - As this is a general purpose rule, legitimate usage of
      the encode functionality will trigger some false positives.
level: medium
\end{lstlisting}
\end{listing}

(1)  id --- a UUID assigned once and never changed. Lineages are joined on it rather than on file path, because paths move under rename while the UUID does not.

(2)  tags --- self-reported ATT\&CK labels, discussed in Section 2.3. attack.defense-evasion names a tactic, what the adversary is trying to achieve \citep{strom2020}; attack.t1027 names a technique, Obfuscated Files or Information. Both are shown as they stood at the snapshot. Four weeks later SigmaHQ re-tagged the whole corpus to ATT\&CK v19, under which this tactic is renamed Stealth and a further tactic, Defense Impairment, is separated from it. Section 2.3 returns to what that implies for an analysis that spans nine years.

(3)  logsource --- names the event stream the rule reads, and so bounds what the rule can ever see.

(4)  detection --- the object of this study. Every quantity reported in Section 5 derives from changes to this block.

Three conventions govern how a detection block becomes a boolean expression, and one consequence of them carries the rest of the paper. Keys within a block are conjoined; values under a single key are disjoined; and a block written as a list of maps is likewise disjoined, so that the rule of Listing 1 fires on (P1 OR P2) AND P3. The consequence is that a list-valued field is a single predicate holding many literals, not many predicates. Image|endswith: ['\textbackslash{}a.exe', '\textbackslash{}b.exe', '\textbackslash{}c.exe'] is one predicate. Appending a fourth entry changes the set of records the rule matches while leaving the number of predicates, and the shape of the condition, exactly as they were. The structural size of a rule and the set of events it matches are therefore distinct quantities that can move independently, and the asymmetry follows directly from conjunction across keys and disjunction within lists. Section 4.2 and Section 5.2 both depend on it.

A second property makes narrowing measurable from rule text alone. Sigma rules are evaluated per record: there is no correlation, thresholding or sequencing across records, and no state is carried between events. Whether a given record raises an alert is fully determined by the rule as written. A change in what a rule declines to alert on is therefore a change in the rule's text, and can be recovered statically from the repository's revision history without reference to any deployment.

\subsection{How suppression is expressed}

Section 1.1 set out why a rule comes to be narrowed. The maintainer of a noisy rule has four courses open and three of them are unavailable or worse: removing the benign generator from the environment is rarely within their authority; rewriting the rule more precisely is expensive, requires revalidation and is frequently impossible because the record contains nothing that separates the two cases; lowering the severity leaves the volume unchanged; and deleting the rule surrenders the coverage entirely. Recording an exception for the specific benign generator is cheap, is targeted, and preserves the rule. It is the rational choice, it is the one practitioners describe themselves making \citep{vermeer2023}, and it is the one the corpus shows them making. Nothing in this paper argues otherwise about any individual decision; the claim advanced is about what becomes of the exceptions in aggregate once they are in place, and about the absence of any process that revisits them.

One term needs fixing before the mechanism is described. In operational usage, suppression often refers to platform-side handling of alerts --- throttling, deduplication, or the automated prioritisation and grouping of alerts after a rule has fired \citep{vanede2022,wang2024}. That is not the sense used here. Throughout this paper, suppression means a change to the content of the rule itself, recorded in the repository, that narrows the set of records the rule will match. It is a property of rule text, and it is therefore visible in version history.

Sigma expresses an exception as a negation. A named block is introduced alongside the existing selections, and the condition becomes a conjunction with that block's complement. Listing 2 is the entire mechanism in four lines.

\begin{listing}[htbp]
\caption{A predicate-level exclusion. Rule win\_security\_susp\_scheduled\_task\_delete.yml (UUID 7595ba94-cf3b-4471-aa03-4f6baa9e5fad) at commit 356ab98a, 9 December 2022, committed by Florian Roth with the message ``fix: FPs with Important Scheduled Task Deleted''. Unmarked lines are unchanged context. Four TaskName entries are elided at the mark, and the author's trailing comment is reflowed onto two lines; nothing else is altered.}
\label{listing:2}
\begin{lstlisting}
  detection:
      selection:
          EventID:
              - 4699                          # Task Deleted Event
              - 4701                          # Task Disabled Event
          TaskName|contains:
              - '\Windows\SystemRestore\SR'
              - '\Windows\Windows Defender\'
              - '\Windows\BitLocker'
              [... four further entries, unchanged ...]
+     filter:
+         SubjectUserName|endswith: '$'
+         # False positives during upgrades of Defender, where its
+         # tasks get removed and added
-     condition: selection
+     condition: selection and not filter
\end{lstlisting}
\end{listing}

The selection is untouched: the rule still asks precisely the question it asked before, over the same seven task names. What has changed is that it now declines to alert in one further circumstance --- when the subject user's name ends in \$, which is to say when the acting account is a machine account rather than a person. The maintainer's reason is recorded in the rule itself, as a trailing comment: Defender's scheduled tasks are removed and re-added during upgrades, and each upgrade was raising an alert. The reasoning is sound and the fix is proportionate. It is also permanent by default. Neither the rule nor the repository records any condition under which it should be reconsidered.

There is a second way to write the same exception, and it is the reason for the methodological claim in Section 5.2. Where a block of exclusions already exists, a further exclusion can be added by appending a value to a list that block already holds. Listing 3 shows one, together with the whole of the change made to the file.

\begin{listing}[htbp]
\caption{A value-level exclusion --- the complete file diff. Rule win\_susp\_proc\_access\_lsass.yml (UUID a18dd26b-6450-46de-8c91-9659150cf088) at commit 8f22165f, 28 November 2021, committed by Florian Roth with the message ``fix: FPs noticed with Aurora''. One line was added and nothing else in the file changed --- not even the modified date. Unmarked lines are unchanged context, reproduced from the same revision; elisions are marked.}
\label{listing:3}
\begin{lstlisting}
  detection:
      selection:
          TargetImage|endswith: '\lsass.exe'
          GrantedAccess|endswith:
              [... 24 access-mask suffixes ...]
      # Absolute paths to programs that cause false positives
      filter1:
          SourceImage:
              - 'C:\WINDOWS\system32\taskmgr.exe'
              - 'C:\Program Files\Malwarebytes\Anti-Malware\MBAMService.exe'
              - 'C:\PROGRAMDATA\MALWAREBYTES\MBAMSERVICE\ctlrupdate\mbupdatr.exe'
+             - 'C:\WINDOWS\system32\taskhostw.exe'
      [... filter2 to filter6 and filter_generic, unchanged ...]
      condition: selection and not filter1 and not filter2 and not
                 filter3 and not filter4 and not filter5 and not
                 filter6 and not filter_generic
\end{lstlisting}
\end{listing}

Nothing structural has moved. The file still contains the same seven exclusion blocks; the condition line is untouched; the number of predicates is unchanged, because --- as Section 2.1 established --- SourceImage holding a list of values is one predicate containing several literals. Only the list grew, from three literals to four. The rule's behaviour nonetheless changed: process access to LSASS memory originating from taskhostw.exe no longer alerts, on a rule whose purpose is to detect credential dumping. A method that compares rule structure between versions, as the prior longitudinal analysis does \citep{long2026}, cannot register this edit as a narrowing at all. Section 4.2 therefore defines exclusion growth at both predicate and value level, and Section 5.2 reports what share of all suppression the second form accounts for.

A third property of the format bears on how readily exclusions accumulate. Where the exclusion blocks are named individually the condition must name each of them, so that every further exclusion requires an edit to that line; the LSASS rule of Listing 3 names seven. Mature rules commonly adopt wildcard grouping instead --- a condition reading not 1 of filter\_main\_* rather than naming particular blocks --- after which a further exclusion is added by writing a new block and doing nothing else. The cost of adding is thereby reduced to very nearly zero. The cost of removing is not reduced correspondingly, because removal requires knowing what the exclusion was for, whether the circumstance that prompted it still obtains, and whether anything since has come to depend on it. Adding is somebody's task at the moment the noise appears. Removing is nobody's. Section 5 reports the consequence.

SigmaHQ has since introduced standalone Sigma filter rules: separate objects that apply exclusions centrally across many detection rules at once. Filters were added, together with correlation rules, in Sigma specification v2.0 in August 2024 \citep{sigmahq2024}. Two consequences follow. Exclusions expressed in that form may not appear within the rule lineages analysed here, which is recorded as a limitation in Section 4.1; since the mechanism existed for roughly the final twenty months of the observation window, that gap is bounded rather than open-ended. And the community's decision to formalise exclusions as first-class objects is independent evidence that exclusion management had become a recognised problem in practice.

This section has described how an exclusion is expressed. How an exclusion is detected --- and distinguished from relaxation, from refactoring, and from narrowing accompanied by compensating growth in coverage --- is the subject of Section 4.2.

\subsection{ATT\&CK tagging and the corpus}

MITRE ATT\&CK is a catalogue of adversary behaviour organised into tactics --- what an adversary is trying to achieve --- and techniques, the specific method by which it is achieved \citep{strom2020}. A technique may be divided into sub-techniques. Sigma rules label themselves with both, as entries in the tags list; the rule of Listing 1 carries attack.defense-evasion and attack.t1027.

These labels are what make it possible to ask whether a suppression matters. A rule is not interesting in itself; what matters is the adversary behaviour it provides coverage for, and whether any other rule covers the same behaviour. The tags supply that join. They convert a local edit --- one line added to one file --- into a statement about coverage of a named technique, and they allow the corpus to be asked whether a given narrowing is cushioned. It frequently is not: the corpus covers 329 techniques with a median of three rules each, but 84 of them, a quarter of the total, are covered by exactly one rule. Section 4.4 describes the join and Section 5.4 reports what the answer turns out to be.

Mechanically, the tags field is an unvalidated list of strings. Convention reserves the attack. prefix for references to ATT\&CK --- a tactic in lower case with words hyphenated, a technique as t followed by four digits and an optional three-digit sub-technique, and software and group identifiers as s and g followed by four digits --- while the prefixes car., cve., detection. and stp. carry references to other catalogues. Nothing enforces any of this at the point a rule is written, and the consequences are visible in the corpus: one rule tags attack.ds0005, a data-source identifier written where a tactic belongs, which names neither a tactic nor a technique and is the plainest available evidence that the field is unchecked. The pre-2021 underscore spelling of tactic names survives at the snapshot only in deprecated rules, on 116 tag instances; the live rule set has been brought fully onto the hyphenated form, which is worth noting for the contrast --- when this project decides a convention needs correcting, it corrects it across the whole rule set.

Three properties of the tags therefore bound what the analysis can claim, and Section 7 returns to each.

First, accuracy. Tags are self-reported, and no process checks them against what the rule actually matches. The rule win\_scm\_database\_handle\_failure.yml, which detects failed attempts to obtain a handle to the service control manager database, is tagged attack.t1010, Application Window Discovery --- a different behaviour entirely.

Second, granularity. Tagging is incomplete rather than uniform. Of the 3,110 rules in the main rule set at the snapshot, 348 --- 11.2 per cent --- carry a tactic but no technique at all, and 1,058, some 34 per cent, carry only parent techniques in cases where sub-techniques exist. Rules written before sub-techniques were introduced acquired them later, if at all.

Third, non-stationarity. The catalogue itself moves under the corpus. Four weeks after the snapshot, SigmaHQ re-tagged its rules to ATT\&CK v19 in a single commit (34c5d66c, 29 April 2026), under which the tactic Defense Evasion is renamed Stealth and a new tactic, Defense Impairment, is separated from it. A coverage analysis keyed on tactic names would not survive that commit intact.

Two consequences follow for the method. Coverage is keyed on technique identifiers rather than on tactic names, because identifiers survive renamings that names do not. And coverage is assessed as of the snapshot while suppression dates are historical, so a rule's coverage set is treated as fixed even though its tags may have changed during the observation window. Neither choice is without cost, and Section 7 states the residual risk in each. Tag accuracy itself is not assumed: a sample is checked by hand against the behaviour each rule matches, and the observed error rate is reported rather than taken on trust.

The corpus is the SigmaHQ repository, analysed through the prepared lineage bundle released with the prior longitudinal study \citep{long2026}. A lineage is the version history of a single rule, identified by its UUID; a revision step is an adjacent pair of versions within one lineage, and is the unit of analysis throughout. At the snapshot, roughly one rule in five carries an exclusion: of the 3,110 rules in the main rule set, 653 contain at least one filter block and 671 have a negation in the condition. A single snapshot understates the extent of exclusion, since exclusions are added, modified and occasionally removed across the history. Section 4.1 describes the corpus in full, and Table 1 gives the descriptive counts.

\section{Related work}
\label{sec:related}

Three literatures bear on this study, and none of them answers its question. The first describes how detection rules are tuned in practice, and does so from the accounts of the people who tune them. The second measures what rules and alert streams look like at scale. The third asks whether the detections that result hold up against an adversary. This section takes each in turn.

\subsection{How detection rules are tuned in practice}

The practice of narrowing a noisy rule has been studied through interviews and fieldwork rather than through the artefacts the practice leaves behind \citep{sundaramurthy2015,kokulu2019,alahmadi2022,vermeer2023,teuwen2025}.

\citet{sundaramurthy2015} embedded researchers trained in anthropological method inside security operations centres as working analysts, and account for analyst burnout as a self-reinforcing cycle of human, technical and managerial factors. Their contribution here is to establish attention as the binding constraint in security operations. That constraint is what makes narrowing a noisy rule the rational response rather than a lazy one, and this paper takes it as given: the question is not why exclusions are added but what the accumulation of them amounts to.

\citet{kokulu2019} interviewed managers and analysts in the same organisations and found the two groups giving contradictory accounts of their own operation. The implication for this study is specific. If the people inside a security operation do not share an accurate picture of how it works, a property that emerges only across years of small decisions is exactly the kind of thing none of them would see. A ratchet is invisible from inside it.

\citet{alahmadi2022} report analysts describing false-positive rates approaching 99 per cent and the coping behaviour that follows. Their account establishes the pressure from the analyst's side of the queue. It does not follow that pressure into the rule text, which is where it leaves its record.

\citet{vermeer2023} is the closest qualitative counterpart to this paper. From interviews with seventeen professionals at managed security service providers, they describe how network intrusion detection rules are created, acquired and managed, and characterise tuning as an ongoing and largely undocumented negotiation between supplied rules and local conditions. What they establish from practitioner accounts --- that exclusions are added routinely and revisited rarely --- this paper measures in the artefacts of one public corpus.

\citet{teuwen2025} approach the same problem from the other end, deriving principles for writing rules that generate less noise in the first place. Their work is the design-time counterpart to the criterion offered in Section 6.5: they ask how to write a rule that will not need narrowing, and this paper asks what to do about narrowing that has already happened and has stood for years.

All five report what practitioners say about tuning, and they agree with one another. None of them measures what the rules record.

\subsection{The alert stream, and the responses to it}

A second body of work measures the alert stream itself and proposes ways of coping with it, establishing why the noise is structural, what a real queue contains, and what the field has done about it \citep{axelsson2000,vanede2022,wang2024,yang2024}.

\citet{axelsson2000} established the arithmetic that makes the problem structural rather than a matter of rule quality: where the base rate of intrusion is very low, even a detector with excellent per-record accuracy produces an alert stream that is overwhelmingly false. Everything that follows in this literature is a response to that result.

\citet{yang2024} supply the measurement. Across a real security operations centre they observe between 24,000 and 134,000 alerts per day, of which 0.01 per cent correspond to true attacks, 27 per cent to attack attempts that did not lead to compromise, and 49 per cent to correct matches on activity that turned out to have a business-justified explanation. That last category is the population this paper's subject matter is written for: an exclusion exists to stop a rule alerting on precisely such activity. Their work measures the input to the process studied here; it does not follow what maintainers then do to the rules.

The dominant response in the literature has been to make the alert stream more tractable downstream rather than to change the rule. \citet{vanede2022} apply semi-supervised contextual analysis to group and score security events so that analysts inspect far fewer of them, and \citet{wang2024} prioritise alerts by reinforcement learning over multi-step attack context. Both are effective and both leave the rule untouched.

That is the division this paper sits across. Triage acts on the alerts a rule produces; narrowing acts on the rule itself, permanently, and is what practitioners in Section 3.1 describe themselves doing. The first has an extensive literature. The second has almost none.

\subsection{What the detection artefacts themselves record}

A third strand reads the detection artefacts themselves --- their coverage claims, their quality, their maintenance, and what an adversary can do with them \citep{strom2020,decan2018,vermeer2022,uetz2024,virkud2024,long2026,tyagi2026,dectot2026}.

\citet{strom2020} set out the design and intended use of MITRE ATT\&CK, which supplies the vocabulary in which detection coverage is now routinely expressed and in which the null result of Section 5.4 is stated. \citet{virkud2024} examine how that vocabulary is used in practice, evaluating four ATT\&CK-annotated rulesets --- three commercial and the crowdsourced Sigma corpus --- for commonalities, under-utilised regions of the matrix, and consistency of labelling across products detecting the same threats. Their finding that labelling is not consistent between products is independent support for the position taken in Section 2.3: ATT\&CK tags are self-reported and must be treated as such.

\citet{vermeer2022} measure evolution at the level of the ruleset rather than the rule, following 130,000 network intrusion detection rules across hundreds of monitored networks at a managed security service provider, and find rulesets growing steadily with almost no overlap between the sets different organisations arrive at. Their unit is the presence or absence of a rule; the unit here is the content of one rule across its own revisions. The two are independent, and can move in opposite directions: a ruleset may grow while every rule within it is narrowing. \citet{long2026} provide the only longitudinal account of how the content of these rules changes. Its scope and the boundary it draws are set out in Section 1.2 and are the starting point for this work. \citet{tyagi2026} assesses the same Sigma corpus for quality across six dimensions, including false-positive risk and taxonomy correctness. The two differ from this paper in the same way: one classifies change by syntactic transformation, the other assesses quality at a single instant, and neither measures what a rule stops matching over time.

Repository mining of detection content is itself an emerging method. \citet{dectot2026} analyse 8.4 million YARA rules across 1,853 repositories, combining repository mining, static quality assessment and dynamic benchmarking, and report a median repository inactivity of 782 days and a median technical lag of over four years. Their finding is the complement of this one: detection content, once published, is frequently not maintained at all, whereas the corpus examined here is actively maintained and the narrowing accumulates anyway.

The pattern is not peculiar to security. \citet{decan2018} measure technical lag across the npm package ecosystem and find dependencies accumulating out of date faster than they are updated, in collections that are under active maintenance. Accumulation without revision appears to be a property of maintained artefact collections generally, and the contribution of this paper is to measure it where the artefact is a detection and the accumulation costs coverage.

Finally, \citet{uetz2024} establish that the gaps such rules contain are exploitable rather than theoretical, constructing evasions for widely deployed SIEM rules and defeating nearly half of them in a live enterprise network. Their work treats the ruleset as a fixed object and asks what an adversary can do with it. This paper asks how the object came to have the shape it has.

Three literatures bear on this study and none of them answers its question. Work on security operations establishes, from interviews and fieldwork, that detection rules are narrowed under pressure and rarely revisited, but it reports what practitioners say rather than what the artefacts record. Work measuring alerts and rulesets establishes the scale of the noise problem and the shape of rule change over time, but the one longitudinal study among them classifies revisions by syntactic transformation, which cannot observe narrowing expressed through negation or through the addition of a value to an existing list. Work on coverage and evasion establishes that detections can be defeated, but treats the ruleset as a fixed object rather than as something that changes under maintenance. This paper measures what the first literature describes, using a method the second lacks, to quantify a consequence the third assumes.

\section{Method}
\label{sec:method}

Section 2.1 established that a Sigma rule is evaluated against one log record at a time, with no state carried between events. What a rule declines to alert on is therefore fully determined by its text. That is the licence for everything in this section: suppression is measured from the repository alone, without reference to any deployment. What follows defines the unit of analysis, the test applied to it, the three narrowing mechanisms that test cannot see, how the resulting exclusions are characterised and followed, and how the whole procedure was checked against hand labelling.

\subsection{Corpus and reproduction}
\label{sec:corpus}

Enterprise detection rules are among the most closely held configuration a security team owns. They disclose both what an organisation can see and what it cannot, and they are not released to researchers; studies of analyst practice have had to proceed by interview for this reason \citep{alahmadi2022}. SigmaHQ is the exception. It is a public repository of detection rules to which practitioners from many organisations contribute, and because it is kept under version control, every change made across nine years is recorded together with its date, its author, and whatever reason its author chose to give. It is the only setting in which the decisions studied here can be observed at all.

What SigmaHQ is not should be said at the outset rather than conceded later. It is not an enterprise security operations centre. Its maintainers respond to reports arriving from many environments rather than tuning against one, and a finding about SigmaHQ is therefore not directly a finding about any particular deployment. It is something else, and arguably of wider consequence. SigmaHQ sits upstream of a large number of commercial and open detection pipelines, and an exclusion committed once is inherited by every consumer that pulls the ruleset. Few of them read the diff. A decision that is locally correct in the environment which prompted it becomes, downstream, a blind spot in environments that never ran the software concerned. Section 7 sets out what this does and does not license.

The corpus is analysed through the prepared lineage bundle released by \citet{long2026}, which aligns rule versions to one another across renames, relocations, splits and merges. Tracking rules by file path instead would misrepresent the corpus badly, since a rename would appear as one rule vanishing and another being created --- one of the documented hazards of mining repository history \citep{kalliamvakou2016}. The notions of a lineage and of a revision step are those defined in Section 2.3, taken from that work rather than re-derived here. The bundle was captured on 10 April 2026; the latest revision it contains is dated 1 April 2026. That capture is what ``the snapshot'' denotes throughout this paper, and every duration reported in Section 5 is right-censored at the capture date. Table 1 describes the corpus.

\begin{table}[htbp]
\centering
\small
\caption{The Sigma corpus as analysed. Counts are taken at the bundle capture date. The study population is the 2,355 lineages carrying at least one predicate-changing revision, and is the denominator for every proportion reported in this paper.}
\label{table:1}
\begin{tabular}{@{}p{0.34\linewidth}p{0.60\linewidth}@{}}
\toprule
Quantity & Value \\
\midrule
Rule files in the main rule set & 3,110 \\
Lineages, all & 4,204 \\
Lineages with a predicate-changing revision (study population) & 2,355 \\
Predicate-changing revisions (unit of analysis) & 8,234 \\
Revision steps per lineage & median 2 $\cdot$ mean 3.5 $\cdot$ 830 have only one \\
Earliest version in the corpus & 27 December 2016 \\
Latest revision in the bundle & 1 April 2026 \\
Bundle capture date (censoring horizon) & 10 April 2026 \\
Observation window & 9.3 years \\
Lineages carrying ATT\&CK tags & 2,127 (90.3\%) \\
ATT\&CK techniques covered & 329 \\
--- covered by exactly one rule & 84 (25.5\%) \\
--- covered by two to four rules & 115 \\
--- covered by five or more rules & 130 \\
Rules per technique & median 3 $\cdot$ max 173 \\
Suppressions per lineage & mean 0.70 $\cdot$ median 0 $\cdot$ max 32 \\
\bottomrule
\end{tabular}
\end{table}

Note. 830 lineages in the study population have a single revision step and therefore no opportunity to narrow. Proportions of lineages that never narrowed are consequently statements about editing activity as much as about narrowing; Section 5.1 conditions on lineages whose exclusion set was actually modified.

Of 4,204 Sigma lineages, 2,355 contain at least one revision that alters the rule's predicate logic. That subset is the study population and the denominator for every proportion reported here, and the proportion it represents agrees with the figure of approximately 56 per cent reported by \citet{long2026}. The lineages excluded were either created and never meaningfully revised, or touched only for formatting, spelling or metadata; a rule whose detection logic never changed carries no information about how detection logic changes.

Everything measured in this paper is built on the pipeline released with the prior work. That pipeline identifies which revisions changed detection logic, aligns versions to one another, and emits the structural records this study reclassifies. An error anywhere in it would be inherited silently and would not announce itself in any result. Published computational artefacts frequently fail to re-run at all \citep{collberg2016}, so nothing here was taken on trust. It was therefore run against its own released data, and its published figures checked, before any extension was attempted. Table 5 of \citet{long2026}, which characterises how rules evolve structurally over their lifetimes, reproduced exactly across all five categories in both of their corpora. Table 6, which characterises reversions, reproduced exactly as well: 216 Sigma lineages with at least one reversion, 383 A--B--A triplets, and a median restoration time of 114.5 hours.

Reproducing that second table is also what made the present question tractable. A restoration time is conditioned on restoration. It describes revisions that were later undone and is silent, by construction, about revisions that were not. The 114.5-hour median therefore characterises the fastest-moving part of rule curation and cannot speak to the part that does not move at all. What was required was a measure that does not condition on the outcome: one that starts when an exclusion is added and follows it whether or not it is ever removed. Sections 4.2 and 4.6 construct that measure.

Four discrepancies were found during reproduction, reported to the authors, and are set out in Appendix A. None of them affects the conclusions of the prior work. One bears directly on the present analysis and is therefore recorded here rather than left to an appendix. The lineage metadata contains 4,358 Sigma lineages where Table I of the prior work reports 4,204, a difference produced by a filter applied before analysis. That same filter causes a version's index within the filtered version list to diverge from its position in the unfiltered list of commits, so that joining the two by position recovers the correct commit only 19.7 per cent of the time. The join used in Section 4.7 is made on commit date or commit hash for this reason. The failure is silent and yields a plausible-looking result, which is the kind of error most worth reporting.

All analysis code is released under an MIT licence at \url{https://github.com/sudaroli1/coverage-decay/}, including the frozen definitions on which Sections 4.2 and 4.4 depend and the derivation record for every figure reported. Upstream data remains under its original licences. The analysis and figure-generation code was written and reviewed with the assistance of Claude (Anthropic). Every figure reported here is produced by a released script from the released data; no figure content is generated by a model. Analyses were run under Python 3.14.6 with pandas 2.3.3, lifelines 0.30.3 and matplotlib 3.10.8.

\subsection{Detecting suppression}
\label{sec:detecting}

A suppression leaves no marker. A Sigma rule carries no field recording that a revision narrowed it, no commit convention identifies one, and the maintainers who make these changes have no reason to label them. The event has to be inferred from what changed between two versions of a rule, and there are 8,234 such changes. Neither can the measure used by the prior work be reused, since a restoration time can only be computed for revisions that were restored. What is required is a test that fires when an exclusion is added, independently of anything that happens to it afterwards.

The test takes one revision step --- an adjacent pair of versions of a single rule --- and returns one of five verdicts: suppression, relaxation, expansion accompanied by a guard, rewrite with no net change, or no change to the exclusion set at all. It is a semantic test rather than a textual one, and it is deterministic: a fixed set of conditions on predicate counts, with no parameters fitted to the corpus and no learned component of any kind. The same revision step always yields the same verdict. Two versions that differ in formatting, in the order of their blocks, or in which branch of the condition an exclusion is written into, express the same detection and must compare as unchanged. What is compared is the set of predicates each version holds and the side of the negation on which each one sits.

Each version is reduced to the canonical predicate form emitted by the prepared bundle \citep{long2026}, in which every field--value test appears as one predicate, marked either as contributing to matching or as held under negation. Three quantities are taken from each version: the number of distinct exclusion predicates, the number of literals contained within those predicates, and the number of selection predicates that actually constrain what the rule matches. Writing these as Ep, Ev and Sp, a revision step is classified by the change in each across the pair.

\begin{quote}\itshape
A revision step is a suppression when the exclusion set grew and coverage did not: when ${\Delta E_{p} > 0}$, or when ${\Delta E_{p} = 0}$ and ${\Delta E_{v} > 0}$; and, in either case, when ${\Delta S_{p} \leq 0}$. A relaxation is the mirror of this: the exclusion set shrank and coverage did not grow.
\end{quote}

Three decisions inside that definition are worth stating, because each was forced by something the corpus contained rather than chosen in advance. Predicates are counted distinct, so that the same exclusion appearing in two branches of a condition tree is one exclusion and not two; 270 of the 8,234 steps contain such duplicates. Growth at the level of literals is counted only where the number of exclusion predicates is unchanged, since where that number moved it is already the signal. And the coverage condition is ${\Delta S_{p} \leq 0}$ rather than ${\Delta S_{p} = 0}$: a revision that narrows coverage while also adding an exclusion is a suppression, not something else.

Two observations forced the refinements that make the definition what it is. The first concerns the coverage condition. In win\_susp\_add\_sid\_history.yml at commit 310e3b7a, a revision added a second event identifier to the rule's coverage and, alongside it, a block named selection3 that appears under negation to stop the new branch firing where the SID history field is empty. The count of negated predicates rises from none to one, so a test asking only whether the exclusion set grew records this as a suppression; in fact the rule now watches an event it could not previously see, and the guard is what makes the new coverage usable. Requiring that the number of constraining selection predicates not increase is what prevents an expansion of this kind from being counted backwards.

A test asking only whether the exclusion set grew would record this as a suppression. It is the opposite. The rule previously watched two event identifiers and now watches three, and the negated block exists solely to stop the new branch firing where the SID history field is empty. Coverage grew and the guard is what makes the new coverage usable. Requiring that the number of constraining selection predicates not increase is what prevents an expansion of this kind from being counted backwards.

The second observation concerns what counts as a constraining selection predicate. A predicate whose only literal is an asterisk, or the empty string, matches every value the field can take and therefore constrains nothing; such predicates arise from how rules are written and converted rather than from any intention to widen coverage. Counted naively they inflate Sp, and a revision that added one alongside a genuine exclusion would be classified as an expansion accompanied by a guard and dropped from the suppression count. Validation item V026 is the clean instance: in a single step a rule gained an exclusion on the parent image path of the Notepad++ updater and, at the same time, a selection predicate on CommandLine whose only literal was an asterisk. After the change the rule matched precisely what it matched before, less the updater. Excluding trivial selections of this kind moved 38 steps out of the expansion-with-guard class, from 111 to 73. The refinement cannot introduce errors in the opposite direction, since a predicate that matches everything genuinely adds no coverage; it removes false negatives only.

Both refinements were found by inspecting the detector's output by hand rather than by reasoning about the definition in advance, and both are among the nine corrections reported with this work. The definitions above are implemented once, in a single module released with the analysis, and are imported rather than restated wherever they are used; divergence between two statements of the same definition in different parts of an analysis was itself the cause of two of those corrections. What this test cannot detect is the subject of Section 4.3, and how far it can be trusted where it does fire is established in Section 4.8.

\subsection{What the detector cannot see}
\label{sec:cannotsee}

A definition of this kind is worth as much as the account of where it fails, and the account is given here rather than in the threats to validity because these are properties of the design rather than doubts about it. Three narrowing mechanisms are invisible to the test set out in Section 4.2, and all three push the measurement in the same direction.

The first is generalisation. A revision may replace several specific exclusions with one broader expression: in lineage\_02634, two named binaries give way to a wildcard over an entire program directory. The number of exclusion predicates falls and the number of literals falls with it, so the test records a relaxation, while the set of records the rule declines to alert on has grown. The second is the compound revision. Validation item V045 withdraws an exclusion on one binary and adds one on another in a single commit; the counts return to where they began and the step is recorded as no net change, although the rule now declines to alert on something it previously caught. The third is case-variant deduplication. Item V085 removes lower-case spellings of three Windows directories while retaining their mixed-case equivalents. Windows paths are not case-sensitive, so nothing changed semantically, but the literal count falls and the step is counted as a relaxation.

Each of these makes the reported ratchet smaller than the truth. Generalisation and compound revision remove suppressions from the numerator; case-variant deduplication adds spurious relaxations to the denominator. The last of these is not corrected, and is left uncorrected deliberately: correcting it would raise the reported ratio, and a known bias that works against the paper's own claim is more useful left in place and declared. The figures reported in Section 5 are therefore floors.

The estimated size of the blind spot is not left to argument. Section 4.8 measures it directly against hand labelling, and 89 per cent of the estimated false negatives fall in the one stratum where the mechanisms described here would put them. The quantitative estimate and the hand inspection agree about where the method is blind, which is the strongest corroboration available for a claim of this sort.

\subsection{Classifying what an exclusion is written on}
\label{sec:classifying}

That a rule has been narrowed does not by itself establish that anything was lost. An exclusion written on an attribute an adversary cannot control costs no coverage in practice, however much it costs in principle: if a rule declines to alert when the acting account is the local system, an attacker who cannot become the local system gains nothing from it. The objection is a serious one and it cannot be answered by counting exclusions. It requires knowing what each exclusion is written on, and whether the value it names can be arranged by someone who wants the rule to stay quiet.

Every exclusion predicate introduced by a suppression is therefore classified twice. The first classification is by attribute: the field the exclusion is written on is assigned to one of six classes --- process, file or registry content; network identity; account identity; system-assigned values; operation parameters; and unclassified --- by pattern match against the field name, with an override table for the fields the patterns do not reach. The override table exists because field naming in the corpus is not consistent, and the classification is reported with no residual: no field falls through to unclassified.

The second classification is the one that carries the argument, and it applies only to fields whose values are filesystem paths. Each literal is assigned an anchor describing what an adversary would have to do to occupy it, taken in order of precedence: a registry path, a user-writable location, a protected system directory, some other absolute path, or a bare filename or suffix. The distinction that matters is between values requiring privileged write to a protected directory and values that do not.

\begin{quote}\itshape
One inference here is load-bearing and is stated rather than assumed. A value beginning with an asterisk derives from a Sigma endswith or contains match, which is location-independent: an exclusion on *\textbackslash{}Windows\textbackslash{}System32\textbackslash{}csrss.exe is satisfied by a file created at C:\textbackslash{}Users\textbackslash{}<user>\textbackslash{}tmp\textbackslash{}Windows\textbackslash{}System32\textbackslash{}csrss.exe, which an unprivileged process can write. Such values are therefore counted as freely enterable. Values of the form *C:\textbackslash{}\dots{} are excluded from that reasoning, since a drive letter cannot appear part-way through a path.
\end{quote}

Anchor alone conflates two properties that need separating. A protected path may name one binary or an entire directory tree, and the second grants an attacker who does escalate everything beneath it. Each literal is therefore also classified by breadth --- whether it names a specific file or a tree or multi-wildcard pattern --- and the two axes are crossed. Content-valued fields such as command lines are held out of this analysis entirely and treated separately, because an adversary writes their own command line and no question of privilege arises; folding them into the path taxonomy in an earlier pass inflated the freely-enterable share and is recorded among the corrections. The classification covers 3,606 field-additions, of which 5,336 literals are path-valued and 1,573 content-valued.

\subsection{Mapping lineages to ATT\&CK techniques}
\label{sec:attack}

Whether a suppression matters depends on what else is watching. An exclusion added to the only rule covering a given adversary technique leaves nothing behind it; the same exclusion on one of a hundred rules covering that technique is cushioned. Making that distinction requires a mapping from rules to the behaviours they claim to detect, and the self-reported ATT\&CK tags described in Section 2.3 are the only such mapping available.

The join is made on the Sigma rule identifier rather than on file path, because paths move under rename and relocation while the identifier does not; joining on path would silently break the mapping for exactly those rules with the longest and most interesting histories. Of the 2,355 lineages in the study population, 2,127 --- 90.3 per cent --- carry at least one technique tag, and between them they cover 329 distinct techniques. A technique covered by exactly one rule in the corpus is treated as sole coverage; one covered by two or more as redundant. The 228 untagged lineages cannot be placed in either group and are excluded from any comparison that depends on the distinction, rather than being assigned a default.

Coverage is assessed as of the snapshot while the suppressions themselves are dated throughout the nine years. This is deliberate and it is the right choice for the question asked, which is whether the exclusions now standing in the corpus sit on rules that have a fallback now. It would be the wrong choice for a question about what the maintainer knew at the time, and Section 7 records the difference.

\subsection{Measuring persistence}
\label{sec:persistence}

The central question is not how often exclusions are added but whether they are ever taken away, and that question cannot be answered by the measure available in the prior work. A restoration time exists only for revisions that were restored. What is needed is a measure that begins when an exclusion is added and continues to follow it whether or not anything further happens, which is the standard situation in survival analysis and is treated as such here.

The clock starts at the commit that adds the exclusion. The exclusion is treated as removed at the first later revision in the same lineage where both the count of distinct exclusion predicates and the count of exclusion literals return to or below their values immediately before the suppression. Both counts are required. A test on predicates alone could never register the removal of a value-level suppression, since that count did not move when the exclusion was added.

Two exclusions from the population are worth stating. Durations shorter than one day are dropped as sequencing artefacts of a single pull request rather than decisions reversed, and 58 of the 1,642 suppressions have no measurable duration at all. That leaves 1,584 observations, of which 203 end in a removal and 1,381 are still in place at the snapshot. The latter are right-censored: they contribute the information that the exclusion lasted at least as long as it was observed, which is precisely the information a naive average of completed durations would discard.

Survival is estimated by the Kaplan--Meier method and read at fixed horizons rather than summarised by a median, because the curve does not reach one half and the median is therefore undefined; any median computed from the duration column would be a median of follow-up time rather than of persistence. Restricted mean survival time over the first three years is reported instead, and median follow-up is estimated by the reverse Kaplan--Meier method so that the flat tail can be shown not to be an artefact of short observation. Where two groups are compared, the comparison is made by log-rank test on the full curves rather than on any single horizon.

One property of the removal test should be declared. It compares counts, not identities, so a later revision that withdraws some other exclusion from the same rule will satisfy it and be recorded as the removal of the exclusion under observation. This can only overstate how often exclusions are removed, so the persistence reported in Section 5.3 is a lower bound on the true figure.

\subsection{Validation}
\label{sec:validation}

A detector that has been refined against the corpus it measures cannot also be trusted on the strength of that refinement. The definitions in Section 4.2 were revised twice in response to output that looked wrong, which is proper practice but leaves the question of how often the final definition is right unanswered. It is answered here by hand labelling.

A sample of 120 revision steps was drawn: 60 that the detector classified as suppressions, and 60 from the four negative strata at 15 apiece. Items were shuffled and presented as predicate differences alone, with commit messages withheld so that the corroboration reported in Section 5.9 remains independent of the labels. A single rater --- the author --- assigned each item to one of three categories: suppression, not a suppression, or unclear. After the detector corrections the arms were recomputed under the final definitions, which moved six items across the boundary and left 66 positive and 54 negative; the items themselves are unchanged, on the view that a human judgement about a diff does not expire when the classifier is corrected. Metrics below are computed on the recomputed arms.

Precision and recall here measure agreement between a fixed rule and a human rater; they are not the performance of a trained model on held-out data, and no part of the corpus is a training set. Precision is the share of decided positives the rater confirmed. Recall is estimated by stratification: the false-negative rate observed within each negative stratum is scaled by the size of that stratum in the full corpus and summed. Because the strata differ enormously in size --- from 73 expansion-with-guard steps to 5,861 with no exclusion change --- the estimate is a weighted extrapolation, and it is reported as a point estimate accompanied by per-stratum intervals rather than as a single confidence interval that would imply more precision than the design supports.

Three disclosures belong with the method rather than with the results. Items the rater could not decide were excluded from both estimates rather than assigned a default, and the rate at which that happened is reported in Section 5.8 because it is itself informative about how hard the boundary is to judge from a diff. The labelling was performed by one rater, so no inter-rater agreement is available; a second rater on a subset remains the single most valuable addition that could be made to this work. And fifteen of the 120 items were discussed with an AI assistant during labelling; those items are flagged in the released labels and the fact is recorded in the declaration on generative AI.

\section{Results}
\label{sec:results}

Seven findings follow. Exclusions accumulate, and roughly a third of the accumulation cannot be seen by comparing rule structure. The accumulation does not reverse: most exclusions are still in place three years after they were added, and whether a rule is the only coverage for an adversary technique makes no difference to whether its exclusions are revisited. The remainder characterise what is being excluded, how the pattern has moved over nine years, and how far the detector can be trusted.

\subsection{Exclusions accumulate}

Across 8,234 revision steps the detector identifies 1,642 suppressions and 304 relaxations. Exclusions are added 5.4 times for every occasion on which one is withdrawn. The remaining steps divide into 5,861 with no change to the exclusion set, 354 rewrites with no net change, and 73 expansions accompanied by a guard.

The ratio is stable under the definitional corrections made during the work, moving only from 1,608 / 299 to 1,642 / 304 across three of them, which is the behaviour one wants from changes that ought to sharpen a definition without moving its substance.

An obvious objection is that the ratchet belongs to a noisy minority of rules rather than to detection rules generally. The 1,642 suppressions come from 524 lineages, and 86 lineages carry five or more. Conditioning on the 559 lineages whose exclusion set was ever modified answers the objection directly: 481 of them, 86.0 per cent, end the observation period net-narrower than they began, against 37 that end net-wider and 41 unchanged. At the level of the rule the ratchet is 13 to 1, stronger than at the level of the revision, because rules that narrow tend to narrow repeatedly while rules that widen do so once.

Two sensitivity checks are worth recording. Excluding the twenty busiest days, which between them account for 326 suppressions, leaves the ratio at 4.9 to 1. And the placeholder rules held in the repository as templates for local customisation, whose edits are arguably not maintenance at all, contribute two suppressions and no relaxations across 17 lineages; the ratio is unchanged.

\begin{figure}[htbp]
\centering
\includegraphics[width=\linewidth]{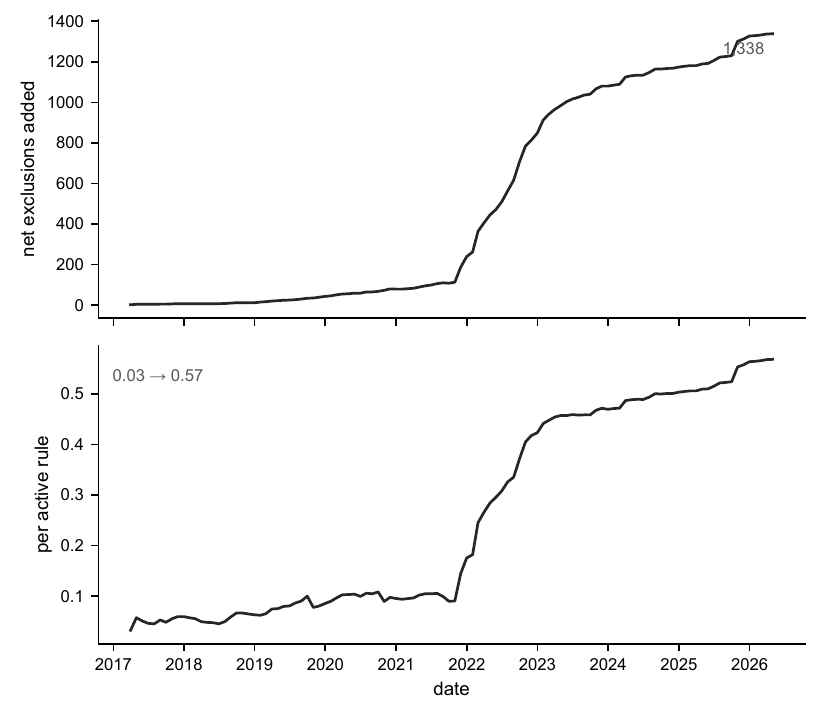}
\caption{Net exclusions added over time. The upper panel gives the raw cumulative count; the lower panel normalises by the number of rules alive at each date and is the one that answers the objection that accumulation merely reflects a growing corpus.}
\label{figure:1}
\end{figure}

\subsection{Two mechanisms, one of them invisible}

Suppression is expressed in two ways, and they are not equally visible. Of the 1,642 suppressions, 1,139 introduce a new exclusion predicate and 503 --- 31 per cent --- add a value to an exclusion that already existed. The second kind leaves the number of predicates, the number of blocks and the condition line all exactly as they were.

The consequence for measurement is direct. A method that compares the structure of a rule between versions observes the first mechanism and is blind to the second, so any structural account of how detection rules evolve undercounts narrowing by roughly a third before the mechanisms of Section 4.3 are considered at all. This is not a criticism of the prior work, whose questions did not require the distinction; it is a statement of what a structural instrument can and cannot see, and it is the reason the detector defined in Section 4.2 counts at both levels.

\subsection{Exclusions are effectively permanent}

Of the 1,584 exclusions whose duration can be measured, 203 are later removed and 1,381 --- 87.2 per cent --- are still in place at the snapshot. Estimated by Kaplan--Meier, 96.6 per cent survive their first month, 93.3 per cent six months, 89.9 per cent one year and 86.7 per cent three years.

No median is reported because none exists: the curve never falls to one half. Restricted mean survival time over the first three years is 978 of 1,095 days, which is to say that an exclusion is in place for about 89 per cent of the three-year window following its introduction. Median follow-up, estimated by reverse Kaplan--Meier, is 1,296 days --- longer than the horizon at which persistence is reported, so the flat tail of the curve is a property of the data rather than of how long the data were observed.

Among the exclusions that are removed, removal happens early or not at all: 52 per cent of removals occur within six months and 76 per cent within a year, and the hazard approaches zero after roughly two years. A mechanism suggests itself, and it is the uncomfortable one. Review of a detection rule is prompted by the rule firing. An exclusion stops the rule firing on the pattern it excludes, and therefore removes the signal that would have prompted its own reconsideration.

\begin{figure}[htbp]
\centering
\includegraphics[width=\linewidth]{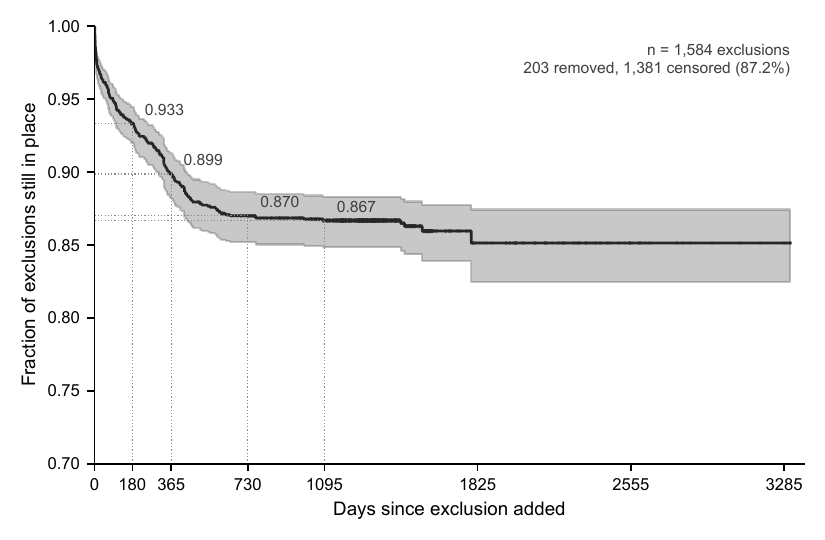}
\caption{Persistence of exclusions, Kaplan--Meier estimate with 95 per cent confidence band (n = 1,584). Vertical marks are censored observations --- exclusions still in place at the snapshot. Survival is annotated at six months, one year, two years and three years.}
\label{figure:2}
\end{figure}

\subsection{Persistence does not depend on coverage redundancy}

If exclusions were being managed with any regard to consequence, one would expect those sitting on the only rule covering an adversary technique to be revisited sooner than those with a dozen rules behind them. They are not. Among the 1,339 exclusions that can be assigned to a group, 71 sit on sole-coverage rules and 1,268 on rules with redundant coverage; 63 of the first group and 1,093 of the second are still in place. A log-rank test over the full curves gives p = 0.49. The two curves are indistinguishable.

This is reported as a finding rather than as a failure to find one. The null has held across three runs on successively larger samples, at p = 0.96, 0.31 and 0.49. What it says is that nothing in the observed process is weighted by consequence: whatever governs the removal of an exclusion, it is not the amount of detection standing behind the rule it was added to. That is consistent with the mechanism proposed in Section 5.3, since a queue driven by alerts firing has no way to know which silences matter.

The 71 sole-coverage exclusions remain interesting, but as a severity argument rather than a persistence one. Those exclusions have no fallback detection behind them, and 63 of them are still in force.

\begin{figure}[htbp]
\centering
\includegraphics[width=\linewidth]{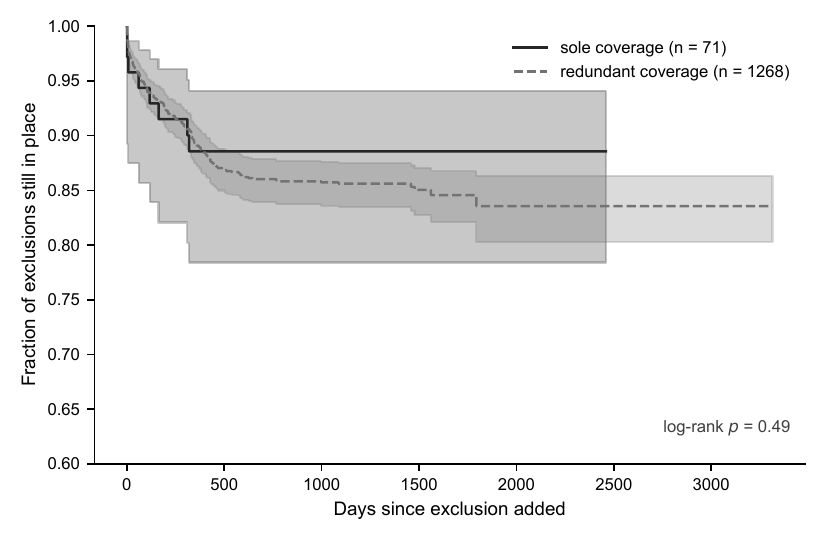}
\caption{Persistence by coverage redundancy. Exclusions on rules that are the sole coverage for an ATT\&CK technique are compared with those on rules whose technique is covered elsewhere in the corpus. The curves are statistically indistinguishable.}
\label{figure:3}
\end{figure}

\subsection{What the exclusions are written on}
\label{sec:written}

The 1,642 suppressions introduce 3,606 exclusion predicates. Classified by attribute, 86.0 per cent are written on process, file or registry content, 10.0 per cent on network identity, and the remainder on account identity, system-assigned values and operation parameters; no field falls outside the taxonomy. On the face of it 98.9 per cent are written on attributes an adversary can influence, but that figure should not be leaned on: when almost everything falls in one bucket, the classification discriminates very little. The question that discriminates is not whether an attribute can be influenced but what influencing it would cost.

Restricting to the 5,336 path-valued literals and classifying each by what an adversary would have to do to occupy it: 58.2 per cent name a bare filename or suffix, 33.0 per cent name a protected system path, 5.9 per cent a user-writable location, and the remainder fall elsewhere. Taken together, 64.1 per cent can be entered by an unprivileged process that chooses a filename, and 33.0 per cent require privileged write access to a protected directory.

\begin{figure*}[htbp]
\centering
\includegraphics[width=\linewidth]{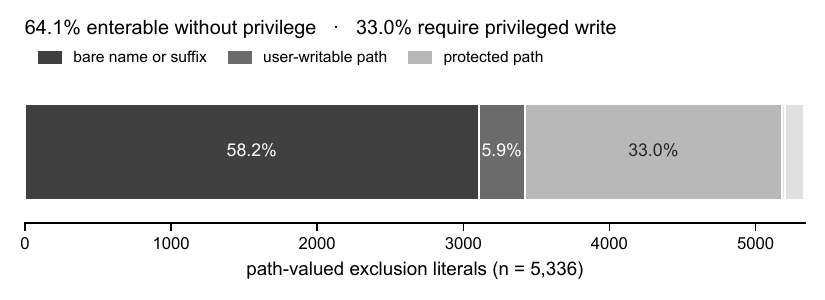}
\caption{Path-valued exclusion literals by what an adversary must do to occupy them (n = 5,336).}
\label{figure:4}
\end{figure*}

The third of exclusions requiring privileged write is a real concession and is made without qualification: for those exclusions, the objection holds. The claim this paper supports is correspondingly narrower than a claim that narrowing removes coverage. It is that roughly two thirds of file-path exclusions can be satisfied by an unprivileged process that chooses a filename.

Anchor is not the whole story, because it conflates difficulty of entry with breadth of grant. A protected path may name a single binary or an entire directory tree, and the second gives an attacker who does escalate everything beneath it. Classifying each literal on both axes: half --- 50.5 per cent --- are free to enter and name a specific file; 12.3 per cent are free to enter and broad; 18.4 per cent require privilege and are narrow; and 13.3 per cent require privilege and are broad.

Two consequences follow, and they cut in opposite directions. The safe harbour is smaller than it appeared: of the 1,759 protected-path exclusions, 708 --- 40.3 per cent --- are directory trees or multi-wildcard patterns rather than named files, which is a blanket amnesty behind a lock rather than a precise carve-out. And the exclusions that are unambiguously costless to the defender are the narrow, privileged ones, which are 18.4 per cent of the total rather than 33.0.

\begin{quote}\itshape
This is also the triage criterion the paper can offer. The 18.4 per cent that are narrow and require privilege are correct engineering and should be left alone. The 12.3 per cent that are broad and freely enterable are where a review should start. The half that lie in between are free to satisfy but buy the attacker a single filename rather than a region.
\end{quote}

Content-valued exclusions --- 1,573 literals on fields such as command lines and registry data --- are treated separately, because an adversary writes their own command line and no question of privilege arises. The question for these is specificity, and they differ sharply by field: command-line exclusions have a median length of 19 characters against 47 for registry and data exclusions, and 20 per cent of them are under eight characters against 1 per cent for registry. Short exclusions are not thereby broad, and the paper does not claim they are: an exclusion of a spaced command form on a rule that detects the unspaced form is short and exactly precise. The distribution is reported as description, not as a finding about danger.

The literals underlying these figures were checked against the raw rule text: 167 of 185 short literals appear verbatim in the source YAML of the same lineage. The check was prompted by a single-character literal that appeared to disable a rule and turned out to be an artefact of the canonical conversion rather than anything in the repository, which is recorded among the limitations.

\subsection{Temporal structure}

Accumulation could be an artefact of a growing corpus. It is not. Net exclusions per rule alive at each date rise from under a tenth of an exclusion per rule in the early years to about 0.57 by 2026, an order-of-magnitude increase in the burden carried by the average rule.

Neither is it an artefact of bulk editing. The twenty busiest days account for 20 per cent of all suppressions, but those days are spread across 2022 to 2025 rather than clustered, the suppressions on them come from 524 distinct lineages, and removing those days entirely leaves the ratio at 4.9 to 1. Nor is the concentration of 58 per cent of suppressions between October 2021 and January 2023 a change in behaviour: total revision activity peaks in the same window, and suppression as a share of all revisions stands at 0.083 within it against 0.058 outside.

Exclusions are not confined to the initial tuning of a new rule. The median rule is 492 days old when an exclusion is added to it; 24 per cent of suppressions occur within 90 days of the rule being written and 44 per cent within a year, so there is a genuine early-tuning spike, but 56 per cent occur after the first year. Taken with the persistence result this gives the lifecycle in a sentence: an exclusion arrives around sixteen months into a rule's life and outlives everything that follows.

\begin{figure}[htbp]
\centering
\includegraphics[width=\linewidth]{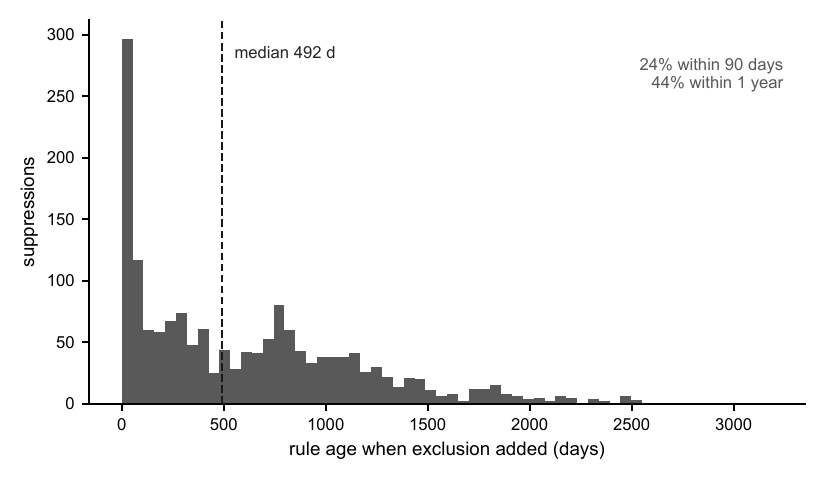}
\caption{Age of the rule at the moment an exclusion is added.}
\label{figure:5}
\end{figure}

Suppression is also a growing share of what maintenance consists of. Across the 77 months carrying at least 50 revisions, suppression rises from 3.3 per cent of all revision activity in the first half of the period to 7.9 per cent in the second. Splitting by mechanism tests a possible mechanical explanation and refutes it: value-level suppression requires an existing list to append to, so its opportunities grow structurally as rules accumulate filters, and that hypothesis predicts the value-level share should carry the trend. It does not. The value-level share moves from 1.5 to 2.2 per cent and is not statistically significant, while the predicate-level share, which has no such dependency, more than triples from 1.7 to 5.7 per cent. The rise is behavioural: as the corpus matures, maintenance shifts from writing detections towards narrowing them.

\begin{quote}\itshape
Months are not independent observations, so the significance of a rank correlation computed over them is optimistic. The effect size is what should be read --- a doubling overall and a 3.4-fold rise for the predicate mechanism --- and serial correlation is recorded as a limitation.
\end{quote}

\begin{figure}[htbp]
\centering
\includegraphics[width=\linewidth]{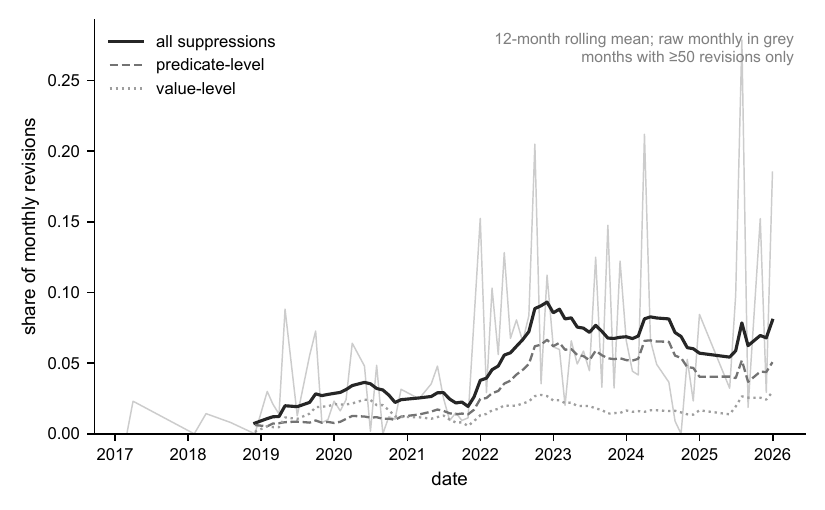}
\caption{Suppression as a share of all revision activity, by mechanism. Raw monthly values are shown faintly behind a twelve-month rolling mean; only months carrying at least 50 revisions are plotted.}
\label{figure:6}
\end{figure}

\subsection{Worked example}

The aggregate results describe a process that is easier to recognise in a single rule. Figure 7 gives the complete exclusion history of image\_load\_dll\_vss\_ps\_susp\_load.yml, which alerts when a process loads vss\_ps.dll, the Volume Shadow Copy Service proxy library. Loading it is characteristic of programs that manipulate shadow copies, which is ordinary behaviour for backup and servicing software and is also what ransomware does immediately before encryption.

\begin{figure*}[htbp]
\centering
\includegraphics[width=\linewidth]{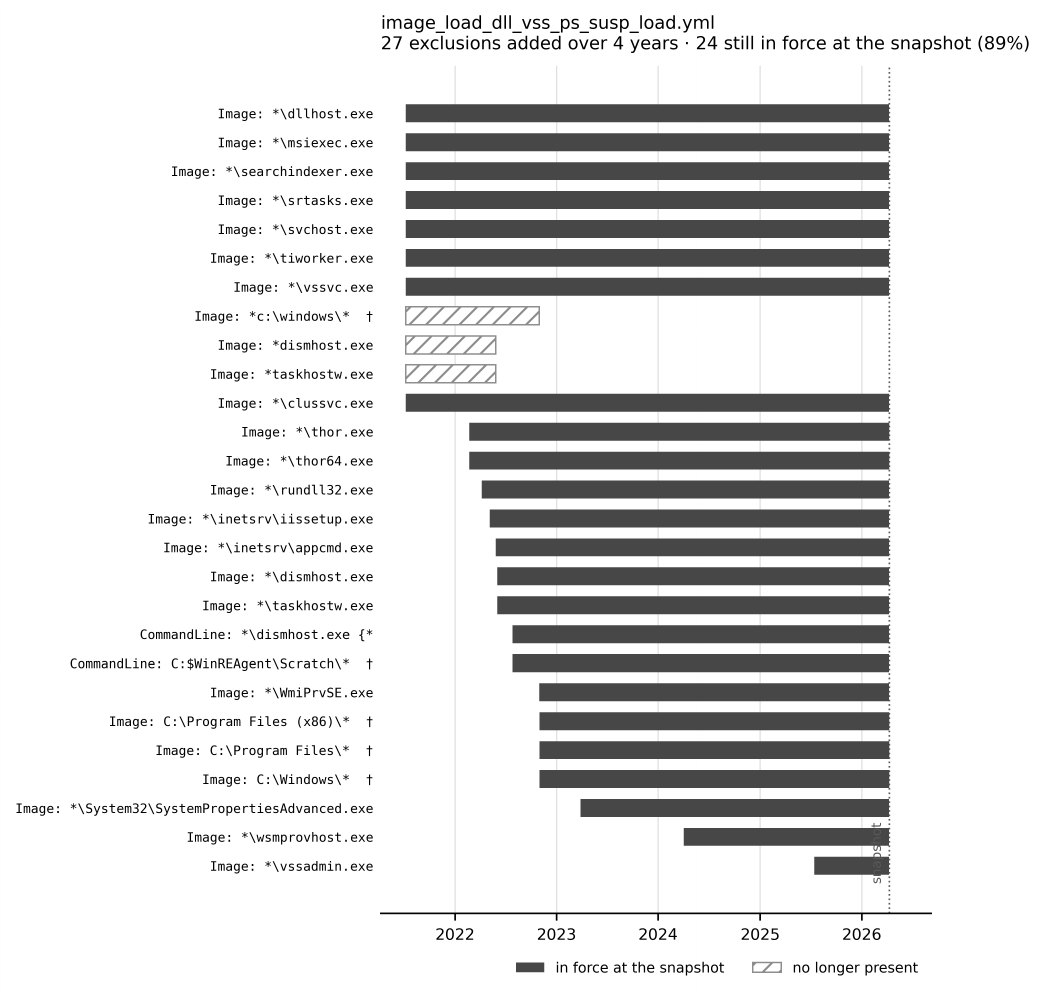}
\caption{The exclusion history of one rule. Each bar runs from the revision that introduced an exclusion to the revision that last carried it, or to the snapshot where it is still in force. Literals marked \dag{} name a directory tree rather than a single file. The exclusions in the principal filter block are conjoined with a prefix requiring the loading process to sit under C:\textbackslash{}Windows\textbackslash{}, which the flattened predicate view does not show; the limitation is the one recorded in Section 5.5.}
\label{figure:7}
\end{figure*}

Twenty-four of the twenty-seven are still in force at the snapshot. More instructive than that number is what happens to the three that are not. None of them is a reconsideration. The exclusions written as dismhost.exe and taskhostw.exe in July 2021 disappear in June 2022 and are replaced the same day by \textbackslash{}dismhost.exe and \textbackslash{}taskhostw.exe --- the identical intention, re-spelled with a leading separator. The third, a contains-match on c:\textbackslash{}windows\textbackslash{}, gives way to a startswith-match on C:\textbackslash{}Windows\textbackslash{}. All three lapses are re-spellings, and this single rule therefore contains, in visible form, the case-variant mechanism that Section 4.3 records as counted against the paper's own hypothesis: the detector sees three relaxations here and none of them relaxes anything.

The breadth of what is excluded does not stay constant either. For the first fifteen months every exclusion names a specific binary. On 31 October 2022 the rule gains three directory trees in a single revision --- C:\textbackslash{}Windows\textbackslash{}, C:\textbackslash{}Program Files\textbackslash{} and C:\textbackslash{}Program Files (x86)\textbackslash{} --- and every executable beneath them ceases to raise this alert. The first sits in the block that already required a Windows prefix and so changes little. The other two are new, and they are the privileged-and-broad case of Section 5.5: an attacker cannot write into Program Files without first escalating, but one who has escalated inherits the whole tree for this detection.

The maintainer is aware of this. The Program Files filter carries a comment in the rule body addressed to whoever deploys it:

\# When using this rule in your environment replace the "Program Files" folder by the exact applications you know use this. Examples would be software such as backup solutions

The exclusion is documented, at the point of writing, as a placeholder that the operator is expected to narrow. It has stood unchanged for three and a half years, and it is inherited unchanged by every deployment that pulls the rule. This is the clearest available statement of the paper's argument, and it is made by the maintainer rather than by us: the individual decision is sound, is taken knowingly, is even annotated with its own remedy, and is nonetheless permanent.

\subsection{Validation results}

Of the 120 hand-labelled items, 58 of the detected arm were decided and 48 of those were confirmed as suppressions, giving a precision of 0.828 with a 95 per cent Wilson interval of [0.711, 0.904]. Estimated by stratification across the negative arms, recall is approximately 0.911; the per-stratum false-negative rates are 1 of 5 for expansion-with-guard, 3 of 9 for rewrites with no net change, and 0 of 13 and 0 of 8 for the no-change and relaxation strata respectively. At that precision the 1,642 detected suppressions imply roughly 1,359 true ones, and the ratio is more robust than either absolute count, since classification error affects numerator and denominator alike.

Two aspects of the exercise are worth reporting for what they say about the phenomenon rather than about the detector. Twenty-seven items --- 22.5 per cent --- could not be decided from the diff at all, which is a high rate and indicates that the boundary between narrowing and rewriting is genuinely hard to judge without knowing the environment the rule runs in. And 89 per cent of the estimated false negatives originate in the rewrite stratum, exactly where the mechanisms described in Section 4.3 predict the method is blind. The quantitative estimate and the hand inspection agree about where the detector fails, which is a stronger claim than either would support alone.

\subsection{Corroboration from commit messages}

The detector reads rule text and nothing else, and commit messages were withheld from the rater during the labelling of Section 4.8, so what maintainers wrote at the time is independent evidence rather than confirmation. Commit subjects are joined to revision steps by lineage and commit date rather than by position, for the reason given in Section 4.1; of the 1,642 suppressions, 1,203 match a unique same-day commit, 439 fall on days carrying several commits and are reported separately, and none is unmatched. Two vocabularies are counted apart, because reason-language --- false positive, noisy, benign, legitimate --- states why an edit was made and nothing in the diff compels it, while mechanism-language --- filter, exclude, whitelist --- merely restates what the diff did. Reason-language appears in 30.9 per cent of the matched suppression commits, in 15.6 per cent of the 186 matched relaxation commits, and in 3.0 per cent of the 4,528 remaining revisions. The enrichment over baseline is 10.3-fold.

Three things follow. The corroboration is not circular: reason-language runs about three times higher than mechanism-language in suppression commits --- 30.9 per cent against 10.8 --- so maintainers are writing that something was a false positive rather than merely restating that they added a filter. The detector separates two kinds of edit it was given no help with: suppressions carry twice the reason-language of relaxations, and the detector never saw a commit message. And the absolute rate is bounded by commit hygiene rather than by the phenomenon, since most Sigma commit subjects are terse; 30.9 per cent is a floor on how often a suppression was prompted by a false alarm, not an estimate of it.

\begin{quote}\itshape
The ratio is the defensible claim and the percentage is not. Reporting 30.9 per cent invites the question of what the other 69 per cent were about, which these data cannot answer; 10.3-fold enrichment over an unrelated baseline can be defended.
\end{quote}

\section{Discussion}
\label{sec:discussion}

\subsection{Why exclusions are not revisited}

The results establish that exclusions accumulate and are not withdrawn. They do not establish why, and the question matters, because a process that fails through inattention needs a different remedy from one that fails through incentive. We propose an explanation that is consistent with the shape of the data and is offered as such rather than as something measured.

Review of a detection rule is prompted by the rule firing. An analyst who sees an alert investigates it, forms a view about whether it was useful, and may act on the rule. An exclusion removes exactly that prompt for the pattern it excludes. The rule continues to fire on everything else and continues to look healthy, while the excluded region generates no evidence at all --- not evidence of absence, but the absence of evidence. A suppression therefore removes the signal that would have prompted its own reconsideration, and the longer it stands the less likely anything is to raise it.

Two features of the data are consistent with this. The removal hazard concentrates sharply and early: 52 per cent of the removals that occur happen within six months and 76 per cent within a year, after which the rate approaches zero. That is the shape one expects if removals are driven by the exclusion being noticed shortly after it was made --- during a review of the same pull request, or by the person who made it --- rather than by any later process. And the coverage-redundancy result is null: exclusions on rules that are the only coverage for a technique are revisited no more often than exclusions with a dozen rules behind them. If review were driven by consequence, the two groups would separate. They do not, because a queue driven by alerts firing has no way to know which silences matter.

That maintainers are aware of the pressure, and are not concealing it, can be read off the artefacts themselves. The Linux network-utilities rule carries this against one of its own detection terms:

- '/telnet' \# could be wget, curl, ssh, many things. basically everything that is able to do network connection. consider fine tuning

The comment sits on a selection term rather than an exclusion, which makes it the more telling: a maintainer recording in the published artefact that a detection is over-broad, and inviting whoever deploys it to narrow it. The invitation is dated September 2024 and still stands. Nothing about the process described here requires anyone to be careless.

\subsection{What this means for the people who run these rules}

The findings reach three audiences, and the middle one has not been addressed at all. Rule maintainers make the decisions and are visible in the record. Researchers measuring rule evolution are affected by one result in particular. Between them sit the organisations that consume the ruleset --- which inherit every exclusion in it, almost never read the diff, and are in the best position to act precisely because they are not the ones who made the decision.

\begin{table*}[htbp]
\centering
\small
\caption{What follows from each finding, and for whom. The middle column names the party in a position to act, which is not always the party whose decision produced the result.}
\label{table:2}
\begin{tabular}{@{}p{0.30\linewidth}p{0.16\linewidth}p{0.46\linewidth}@{}}
\toprule
Finding & Who can act & What follows \\
\midrule
Exclusions are added 5.4 times for every one withdrawn, and 13 times at the level of the rule & Rule maintainers & Addition and removal are not symmetric activities. Only addition has an owner. \\
A third of narrowing is invisible to structural comparison & Researchers, tool builders & Any instrument that compares rule structure is undercounting. Measure at value level as well. \\
86.7 per cent of exclusions are still in force after three years & Ruleset consumers & An exclusion inherited today is, in expectation, permanent. Treat merge as the last review it will receive. \\
Persistence is independent of whether the rule is the sole coverage for a technique & Ruleset maintainers, tool builders & No signal currently distinguishes a costly exclusion from a cheap one. One could be computed. \\
64.1 per cent of path exclusions can be satisfied by an unprivileged process choosing a filename & SOC engineers & Exclusion breadth and anchor are reviewable properties of rule text, before deployment. \\
12.3 per cent are both broad and freely enterable; 18.4 per cent are narrow and privileged & SOC leads & A review queue ordered by these two axes is tractable where a review of every exclusion is not. \\
Suppression rises from 3.3 to 7.9 per cent of all revision activity & The community & Maintenance is shifting from writing detections towards narrowing them. \\
\bottomrule
\end{tabular}
\end{table*}

The practical contribution is the last row but one. An instruction to review your exclusions is not actionable: there are thousands, each was added for a reason, and nothing distinguishes them. Ordering them by two properties of the text --- how broadly the exclusion is drawn, and whether an adversary could occupy it without first acquiring privilege --- makes the task finite. In this corpus that is 657 literals to look at first out of 5,336, with a further 984 that can be left alone on the grounds that they are precisely drawn and cost an attacker something to satisfy. Both properties are computable from the rule before it is deployed.

\subsection{Public discoverability}

These rules are public, and so are the exclusions in them. Anyone may read which conditions a widely deployed detection declines to alert on, and the reading requires no access to any organisation. That is ordinarily a strength of open detection content and it is not being argued otherwise here; the observation is narrower. Where an exclusion is drawn on an attribute that an unprivileged process can choose --- 64.1 per cent of path-valued literals in this corpus --- the circumstance under which a rule stays quiet is both documented and cheap to arrange.

The claim is deliberately limited to what the data support. These windows exist, they persist, and they are discoverable. Nothing here shows that any of them has been used, and this study observes no adversary and no deployment. That such gaps are exploitable in practice is not, however, speculative: \citet{uetz2024} constructed evasions for widely used SIEM rules and found that adversaries could evade nearly half of them in a real enterprise network. Their work establishes feasibility; this one establishes that a particular class of window is opened deliberately, for good reasons, and then left open.

\subsection{Implications for measurement}

One result bears on how this kind of work should be done rather than on what defenders should do. Thirty-one per cent of the narrowing in this corpus is expressed by adding a value to a list that already exists, which leaves the number of predicates, the number of blocks and the condition line unchanged. An instrument that compares the structure of a rule between versions cannot see it. Any measurement of rule evolution built on structural comparison is therefore undercounting narrowing by roughly a third, before the three mechanisms of Section 4.3 are considered at all.

This is not particular to Sigma. It follows from a property shared by most detection-rule formats, and by most configuration languages: a field whose value is a list is one term of the expression however many entries the list holds. Wherever that is true --- Splunk Security Content, Elastic detection rules, YARA string sets, firewall object groups --- the same blind spot exists, and the same correction applies. Measure the set of literals as well as the set of predicates, or accept a known undercount and say so.

\subsection{What a remedy would look like, and what it cannot fix}
\label{sec:remedy}

The results point at three interventions, none of which requires new research and all of which are cheap relative to the alternative of reviewing everything. An exclusion could carry an expiry or review date, so that standing indefinitely becomes a decision rather than a default. The coverage consequence of an exclusion could be surfaced at the moment it is proposed: whether the rule it narrows is the only one covering its technique is computable from the corpus and could be reported in the pull request. And exclusions could be checked for breadth and anchor when they are written, so that a broad, freely enterable exclusion has to be argued for rather than merely merged. The last of these is the smallest and would reach the 12.3 per cent identified in Section 5.5.

That any of this is achievable is not hypothetical, and the corpus contains its own counter-example. In one revision examined during validation, a maintainer replaced five exclusions written on bare filenames --- chrome.exe, firefox.exe and others, each of which an adversary can satisfy by naming a file --- with three absolute paths, dropping two entirely. That is precisely the hardening this section recommends, performed unprompted, by someone who understood the difference between the two forms. The practice exists. What is missing is anything that asks for it.

The boundary of this advice should be stated plainly, because it is narrower than it may appear. What can be identified here is exclusions that are broad, that an adversary could occupy without privilege, and that have stood without review. What cannot be identified is exclusions that are wrong. Whether the thing excluded is genuinely benign depends on the environment in which the rule runs, and no environment is observed in this study. A broad exclusion in an estate where that software genuinely runs everywhere may be the correct engineering decision; a narrow one may be a mistake. The contribution is a way of deciding what to look at first, not a verdict on what is found there.

\section{Threats to validity}
\label{sec:threats}

The substantive limitations of the method are disclosed where they arise, in Sections 4.3, 4.6 and 4.8, on the view that a limitation presented as a design decision is more useful than one presented as a concession. This section summarises them and states the response to each. Three of them bias the results downward, so the ratio reported here is a floor rather than an estimate.

\begin{table}[htbp]
\centering
\small
\caption{Threats to validity and the response to each.}
\label{table:3}
\begin{tabular}{@{}p{0.34\linewidth}p{0.60\linewidth}@{}}
\toprule
Threat & Response \\
\midrule
The corpus is not a deployed configuration & The claim is scoped to the upstream signal reaching organisations that deploy near-stock rulesets. No claim is made about any specific deployment. \\
Not every suppression costs real coverage & Quantified rather than assumed. 64.1 per cent of path-valued literals can be entered by an unprivileged process choosing a filename; 33.0 per cent require privileged write, and for those the objection holds (Section 5.5). \\
ATT\&CK tags are self-reported & Tag quality is spot-checked; at least one mistag was identified by hand. Coverage is keyed on technique identifiers rather than tactic names, which survive the catalogue's own renamings (Section 2.3). \\
Some narrowing mechanisms remain undetected & Value-level suppression is detected and accounts for 31 per cent. Generalisation, compound revisions and case-variant deduplication are not. All three bias the ratio downward (Section 4.3). \\
Right-censoring at the snapshot & Exclusions in force at the snapshot may have been removed since. Survival analysis handles censoring directly, and median follow-up of 1,296 days exceeds the horizon at which persistence is reported. \\
The removal test compares counts, not identities & A later revision withdrawing a different exclusion satisfies it. This can only overstate removals, so 86.7 per cent is a lower bound. \\
Joins to commit metadata & Positional joins recover the correct commit only 19.7 per cent of the time. Joins are made on commit date; the 439 suppressions falling on multi-commit days are reported separately rather than folded in. \\
Literal fidelity to the source YAML & The canonical form is not always faithful to the rule text. 167 of 185 short literals were verified verbatim against the source (Section 5.5). \\
Serial correlation in the trend test & Months are not independent observations, so the rank-correlation significance is optimistic. The effect size is reported alongside it. \\
Errors inherited from the upstream pipeline & The prior work was reproduced before any extension; four discrepancies are documented in Appendix A, none of which affects its headline findings. \\
A single rater in validation & No inter-rater agreement is available, and 22.5 per cent of items could not be decided from the diff. A second rater on a subset is the most valuable remaining addition to this work. \\
\bottomrule
\end{tabular}
\end{table}

\FloatBarrier
\section{Conclusion}
\label{sec:conclusion}

Detection rules narrow over time, and the narrowing does not reverse. Across nine years of the SigmaHQ corpus, exclusions were added 5.4 times for every occasion on which one was withdrawn, and 13 times at the level of the individual rule. Eighty-seven per cent of them are still in force three years later. Whether the rule an exclusion narrows is the only coverage in the corpus for its adversary technique makes no difference to how long that exclusion stands.

A third of this is invisible to the instrument the field has been using. Narrowing expressed by appending a value to an existing list changes no rule structure at all, and any account of rule evolution built on structural comparison undercounts it accordingly. That correction applies wherever detection logic is written with list-valued fields, which is nearly everywhere.

None of this requires anyone to have acted badly, and the record shows that they did not. Every exclusion examined here was a considered response to a real problem. What is missing is not diligence but a signal: nothing tells a maintainer which of their exclusions is broad, cheap for an adversary to satisfy, and standing on the only rule that watches for a given behaviour. Two of those three properties are computable from the rule text before it is deployed, and this paper shows how. What cannot be computed is whether the excluded thing is genuinely benign, which depends on an environment no public corpus contains, and which is where the next study should look.

\subsection{CRediT authorship contribution statement}

Sudaroli Dhananjeyan: Conceptualization, Methodology, Software, Validation, Formal analysis, Investigation, Data curation, Writing -- original draft, Writing -- review \& editing, Visualization.

Kumaran U: Supervision, Writing -- review \& editing.

\subsection{Declaration of competing interest}

The authors declare no known competing financial interests or personal relationships that could have appeared to influence the work reported in this paper.

\subsection{Funding}

This research received no specific grant from any funding agency in the public, commercial, or not-for-profit sectors.

\subsection{Ethics}

This study analyses publicly available software repositories. No human subjects were involved and no personal data were processed.

\subsection{Data availability}

The analysis pipeline, the derived data and the complete validation record are publicly available under an MIT licence. The exact version underlying this paper is archived at \url{https://doi.org/10.5281/zenodo.22107282} (release v1.0.0); development continues at \url{https://github.com/sudaroli1/coverage-decay.} The upstream data are third-party and obtainable as described in that repository: the SigmaHQ rule repository, and the prepared pipeline outputs released with \citet{long2026}.

\subsection{Acknowledgements}

The authors thank E. Long for making the prepared lineage corpus available and for identifying a defect in the structural-operation detector during correspondence, and M. Dacier for the objection that prompted the analysis in Section 5.5 and materially narrowed the claim this paper makes.

\subsection{Declaration of generative AI and AI-assisted technologies in the manuscript preparation process}

The research reported here --- its conception, the formulation of the research question, the design of the method, every methodological decision and correction recorded in Sections 4 and 5, the analysis itself and the interpretation of its results --- is the work of the first author.

During the preparation of this work the first author used Claude (Anthropic; model version and dates of use are recorded in the released repository) for three purposes: assistance in drafting and editing the prose of the manuscript, working from the author's own results, notes and section outlines; assistance in writing and reviewing analysis and figure-generation code; and discussion of 15 of the 120 items in the detector validation exercise (V001--V006, V020--V026, V044 and V045), which are flagged as such in the released label file so that any reader may exclude them and recompute. After using this tool the first author reviewed and edited all content, verified every reported figure against the released data, and the authors take full responsibility for the content of the published article.

No generative tool was used to produce, select or interpret any measurement reported in this paper.

\bibliographystyle{elsarticle-harv}
\bibliography{references}

\end{document}